\documentclass{Interspeech2024}

\interspeechcameraready

\title{Preserving spoken content in voice anonymisation with\\character-level vocoder conditioning}

\name{Michele}{Panariello}
\name{Massimiliano}{Todisco}
\name{Nicholas}{Evans}

\address{EURECOM, Sophia Antipolis, France}
\email{firstname.lastname@eurecom.fr}


\usepackage{tabularray}
\usepackage{ninecolors}
\usepackage{graphicx}
\usepackage{balance}
\newcommand{\utt}{\mathbf{u}}
\newcommand{\x}{\mathbf{x}}
\newcommand{\y}{\mathbf{y}}
\newcommand{\semantic}{\mathbf{s}}
\newcommand{\predicted}{\mathbf{a}}
\newcommand{\ch}{\mathbf{c}}
\newcommand{\scaledict}{\mathbf{w}}
\newcommand{\biasdict}{\mathbf{b}}
\newcommand{\acoustic}{\mathbf{\Tilde{\predicted}}} 

\begin{document}

\maketitle

\begin{abstract}
    

Voice anonymisation can be used to help protect speaker privacy when speech data is shared with untrusted others.
In most practical applications, while the voice identity should be sanitised, other attributes such as the spoken content should be preserved.
There is always a trade-off; all approaches reported thus far sacrifice spoken content for anonymisation performance. 
We report what is, to the best of our knowledge, the first attempt to actively preserve spoken content in voice anonymisation. 
We show how the output of an auxiliary automatic speech recognition model can be used to condition the vocoder module of an anonymisation system using a set of learnable embedding dictionaries in order to preserve spoken content.
Relative to a baseline approach, and for only a modest cost in anonymisation performance, the technique is successful in decreasing the word error rate computed from anonymised utterances by
almost 60\%.

    
\end{abstract}

\section{Introduction}


\emph{Voice anonymisation} consists in the processing of a speech utterance in order to conceal the voice identity of the speaker, while keeping intact the speech content and other paralinguistic attributes.
The study of anonymisation is motivated in part by the need to store, process or share speech data in a manner compliant with privacy regulation such as the GDPR~\cite{literally_the_gdpr_itself, nautsch19c_interspeech}.
The VoicePrivacy initiative~\cite{tomashenko_introducing_2020} has proposed datasets, protocols and metrics to support research in 
voice anonymisation and the benchmarking of candidate solutions.

Voice anonymisation is usually performed using voice conversion to substitute the voice contained within a speech recording with that of a different \emph{pseudo-speaker}.
Anonymisation performance is assessed using an adversarial automatic speaker verification (ASV) model. 
A privacy adversary is assumed to be \emph{semi-informed}, i.e.\ they are aware that the speech data is anonymised and use the same anonymisation system to produce data with which to train an adversarial ASV system, denoted as $\text{ASV}_{\text{eval}}^{\text{anon}}$. The adversary attempts to undermine the anonymisation safeguard by using the $\text{ASV}_{\text{eval}}^{\text{anon}}$ system to reidentify the original speaker in protected speech data. 
The speaker verification equal error rate (EER) achieved by the attacker is estimated using an anonymised dataset. 
The higher the EER, the better the anonymisation.

Some other content must always be preserved - otherwise, the replacement of speech with silence offers a perfect anonymisation solution.  
The 2024 edition of the VoicePrivacy Challenge~\cite{vpc24}
considered the preservation of the spoken and emotional content.
Preservation of the spoken content is evaluated in terms of the word error rate (WER)~\cite{vpc24, vpc2020_results_and_findings, vpc22_perspectives}, derived from the comparison of ground-truth transcriptions to those of anonymised utterances obtained using an automatic speech recognition (ASR) system denoted as $\text{ASR}_{\text{eval}}$.
Preservation of emotional content is evaluated in terms of the unweighted average recall (UAR), derived from the comparison of ground-truth emotion labels to those estimated from anonymised utterances using a speech emotion recognition (SER) system.

Prior work has focused almost exclusively on techniques to better sanitise the voice characteristics of the original speaker.  
Even if the preservation of other attributes is still measured, little-to-no work in the literature has focused on the design of techniques to \emph{expressly} preserve them.
This approach has a fundamentally different emphasis, casting the problem of privacy preservation as that of preserving \emph{desired} attributes, rather than or in addition to sanitising the \emph{undesired} attributes.  
In this paper, we propose a step in this direction.
We report a technique which
still aims
to substitute the voice of an original speaker with that of a pseudo-speaker, but is also designed to 
preserve spoken content.
We do so by conditioning the vocoder which is used for waveform synthesis with the transcription of the original utterance, and training it so that the transcription is unchanged after anonymisation.

We apply our approach to the 2024 VoicePrivacy Challenge~\cite{vpc24} B4 baseline system, which is derived from our previous work~\cite{nac}. While
the system delivers strong 
voice anonymisation, it was found to degrade the spoken content.
Results show that the reported technique is successful in better preserving spoken content, to the point that it outperforms all other baseline systems in terms of the WER, albeit with a modest reduction in voice anonymisation performance. 
While not a design goal, results show that emotion cues are also better preserved.
The proposed conditioning technique is sufficiently generic that it could also be applied to the neural vocoder model of any comparable voice anonymisation system. 


\begin{figure*}[h]
    \centering
    \caption{Overview of the training and inference procedures. Red dashed lines indicate the gradient flow.
    At training time, the system operates in a copy-synthesis fashion, extracting the pseudo-speaker identity from the original input utterance $\utt_i$.
    At inference time, only the green blocks are kept, and the pseudo-speaker is selected at random from the pool of pseudo-speakers as in~\cite{nac}.}
    \includegraphics[width=\textwidth, trim={0 0.5cm 0 0}, clip]{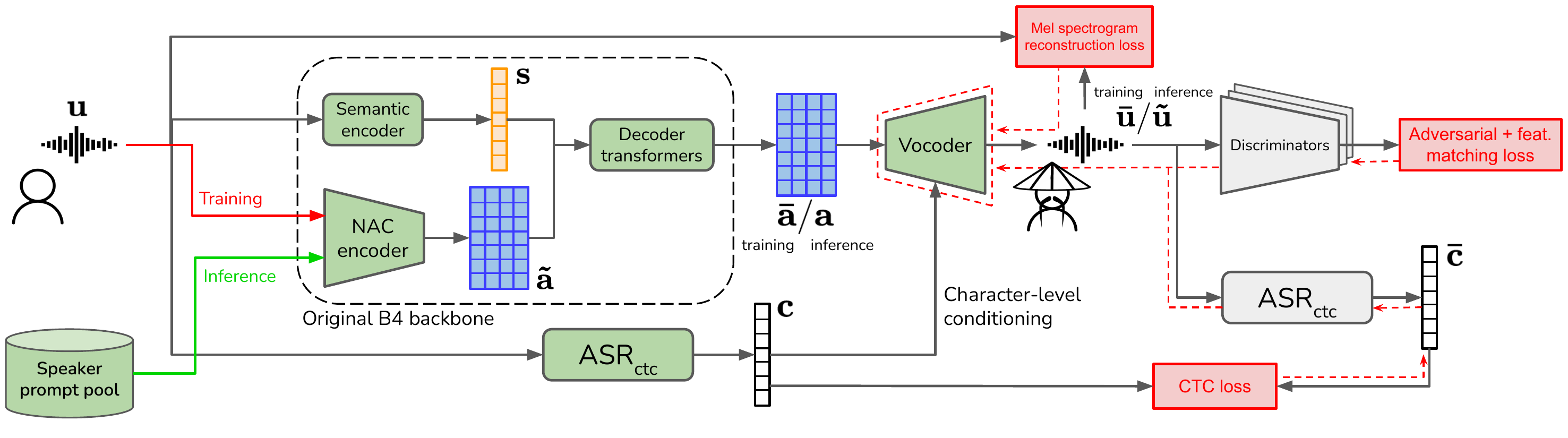}
    \label{fig:train_inf}
\end{figure*}

\section{Related work}
\subsection{Preservation of spoken content}

To the best of our knowledge, few modeling approaches for voice anonymisation were designed to explicitly minimise the WER metric. The closest work in the literature is~\cite{deng_v-cloak_2023}. A connectionist temporal classification (CTC) loss~\cite{ctc} is used during training 
to encourage the preservation of
spoken content. However,
the evaluation method does not align with that of the VoicePrivacy Challenge, most notably in terms of the attack model.
The privacy adversary is uninformed and so the ASV system is not retrained using anonymised data.  This approach is known to result in a substantially weaker attack, and hence gives a potentially unreliable assessment of anonymisation performance.  Other key differences relate to the use of an additional loss term during training, whereas our approach involves a more explicit inductive bias in the model architecture to better preserve spoken content.

Other works have explored a different approach whereby an anonymised speech waveform is synthesised anew from the transcription of the original utterance, therefore erasing the original voice characteristics, and by injecting certain desired attributes into the generated output. The authors of~\cite{T04} used an ASR system to transcribe the input utterance, then employed text-to-speech (TTS) synthesis conditioned on the voice of a pseudo-speaker. This method achieved both a high EER and a low WER but erased the prosody of the input utterance.
Baseline B3 of the VoicePrivacy Challenge 2024, proposed by the same authors in~\cite{meyer_prosody_is_not_identity}, aimed to address this issue by conditioning the TTS model also on pitch and energy values estimated from the input utterance, resulting in a more faithful reproduction of the prosody at the cost of a lower EER.
In this work, we propose a hybrid approach that still relies on voice conversion applied to the input utterance, therefore preserving prosody and paralinguistic attributes, while injecting spoken content information back into the generated output during vocoding.  This encourages the preservation of spoken content with a lower sacrifice in the EER.

We note that other work reported in the context of the 2022 VoicePrivacy Challenge, including two baselines~\cite{vpc22_perspectives, T04, ohnn}, seemingly already shows that spoken content can be preserved to the point that the WER even \emph{improves}.  Decreases in the WER 
might be interpreted to mean that anonymisation  enhances intelligibility of the spoken content, even without specific optimisations to do so.  This is not the correct interpretation. Decreases in the WER
are instead attributed to the 
retraining of the $\text{ASR}_{\text{eval}}$ model using a set of anonymised data~\cite{vpc22_perspectives, vpc2022}, when the anonymisation system itself is trained using a database far larger than that used to train the initial $\text{ASR}_{\text{eval}}$ model.\footnote{Because of the re-training using anonymised data for the 2022 VoicePrivacy Challenge, the ASR system used for evaluation is denoted as $\text{ASR}_{\text{eval}}^{\text{anon}}$~\cite{vpc22_perspectives, vpc2022}
 in related work.}
  In any case, WERs estimated using different ASR systems \emph{and} different evaluation data are not comparable. WER estimates made using the same ASR model are comparable and reveal \emph{increases} in the WER.  
For the 2024 challenge, the 
ASR system is trained only once, and using unprotected data only~\cite{vpc24}. Under this scenario, anonymisation \emph{always} results in a higher WER.
%
We propose a technique to reduce the gap in the WER estimated from unprotected and anonymised data and report its application to one of the 
2024 VoicePrivacy Challenge baselines.

\subsection{Anonymisation using a neural audio codec}
We performed this
study
using the VoicePrivacy baseline B4, introduced in our previous work~\cite{nac}.
As illustrated in the dashed block to the centre-left of Figure~\ref{fig:train_inf},
it combines a neural audio codec (NAC)~\cite{encodec} with a pair of autoregressive decoder-only transformers to apply voice conversion
to an input utterance~$\utt$.
At inference time,
a set of \emph{semantic tokens}~$\semantic$ are first generated using a semantic encoder.
They encode a variety of attributes including the spoken content and prosody in~$\utt$.
The encoder of a NAC is used to extract a set of \emph{acoustic tokens} $\acoustic$, not from $\utt$, but from a recording containing the voice of a pseudo-speaker.
The acoustic tokens are a compressed representation of the waveform containing the pseudo-speaker voice, and comprise two sets: \emph{`coarse'} acoustic tokens which encode longer-term characteristics; \emph{`fine'} acoustic tokens which encode more short-term, fine-grained detail.

Semantic tokens $\semantic$ and acoustic tokens $\acoustic$ are concatenated and fed into the pair of autoregressive transformers, which process the coarse and fine acoustic tokens respectively. The transformers extract the vocal timber of the pseudo-speaker from $\acoustic$ (ignoring other information, e.g. spoken content), and merge it with the semantic content of $\semantic$. The result is encoded in a new set of acoustic tokens $\predicted$.
The NAC decoder then acts as a vocoder, operating upon $\predicted$ to synthesise a waveform output $\Tilde{\utt}$
which then contains speech in the voice of the pseudo-speaker.

While the NAC anonymisation system is competitive with the other state-of-the-art methods in terms of anonymisation performance, it was found to degrade spoken content.
As we show below, 
the vocoder can be conditioned to preserve spoken content with only a modest loss in anonymisation performance.

\section{Proposed method}
The approach is illustrated in Figure~\ref{fig:train_inf} and is built around the NAC-based approach to voice anonymisation described above.
\vspace{-0.1cm}
\subsection{Character-level conditioning}
Informal listening tests reveal that utterances anonymised using the NAC-based approach to anonymisation contain occasional mispronunciations. 
For example, we found cases in which the word `snack' was altered to `stack', or `thick' was altered to `flick'.
We assume that mispronunciations are caused by the vocoder and that they might be tackled 
by training the vocoder to use time-aligned, character-level annotations in addition to acoustic tokens $\predicted$.
To avoid mispronunciation errors,
the vocoder can be conditioned
to use learnable time-aligned, frame-level embeddings which represent the character spoken in the input utterance.
Ideally, embeddings should reflect exclusively spoken content.
So as to avoid privacy-leakage, the voice characteristics should be contained exclusively in the acoustic tokens $\predicted$.


The vocoder model is conditioned at every layer with the output of an auxiliary CTC-based ASR system.
We use the Vocos~\cite{vocos} vocoder architecture, which generates a waveform by processing a set of
NAC acoustic tokens
with a chain of \emph{ConvNeXt} blocks~\cite{convnext} which keep the shape of the intermediate features fixed to $D\times T$ for every layer, where $D$ is the channel dimension and $T$ is the number of time steps.
This property eases the application of a conditioning mechanism to intermediate vocoder layers.
The last layer of the vocoder produces a set of frame-level features which are used by a fully-connected layer to estimate the complex short-term Fourier transform (STFT) of each frame.  A waveform is then synthesised using the inverse STFT.
We employ the version of Vocos designed to operate upon acoustic tokens generated using EnCodec~\cite{encodec}. Since the acoustic tokens $\acoustic$ and hence also $\predicted$ are in the same EnCodec token format, substitution of the NAC decoder (vocoder) with Vocos is straightforward. 

We adjust the vocoder architecture to facilitate its conditioning on
character-level annotations
extracted from the input~$\utt$.
Between each Vocos ConvNeXt layer, we introduce a \emph{character conditioning} layer which we now describe.
As illustrated to the lower left of Figure~\ref{fig:train_inf}, we use a pretrained CTC-based~\cite{ctc} ASR model denoted $\text{ASR}_{\text{ctc}}$ to extract a character sequence $\ch \in \{0, 1 \dots 30\}^{T}$ from the input $\utt$.
Each element of~$\ch$ is an index associated to a set of
CTC characters
including the 26 letters of the English alphabet, the white space character, the apostrophe character, the CTC \emph{null} token, and the \emph{beginning-of-sentence} and \emph{end-of-sentence} tokens.\footnote{The length of $\ch$ could be different to $T$ because of a setting mismatch between $\text{ASR}_{\text{ctc}}$ and the anonymisation system (e.g.\ different hop size). If so, $\ch$ is resized to have length $T$ using nearest-neighbour interpolation.}

\begin{figure}[t]
    \centering
    \caption{Character conditioning layer $k$. Starting from an array of character indexes $\ch$, two matrices $\scaledict_k(\ch)$ and $\biasdict_k(\ch)$ are constructed from embedding dictionaries $\scaledict_k$ and $\biasdict_k$.
    The intermediate vocoder representation $\x_k$ is pointwise-multiplied by $\scaledict_k(\ch)$ then summed to $\biasdict_k(\ch)$. The result $\y_{k+1}$ is passed to the next layer.}
    \includegraphics[width=\linewidth, trim={5cm 0cm 4.55cm 3.2cm}, clip]{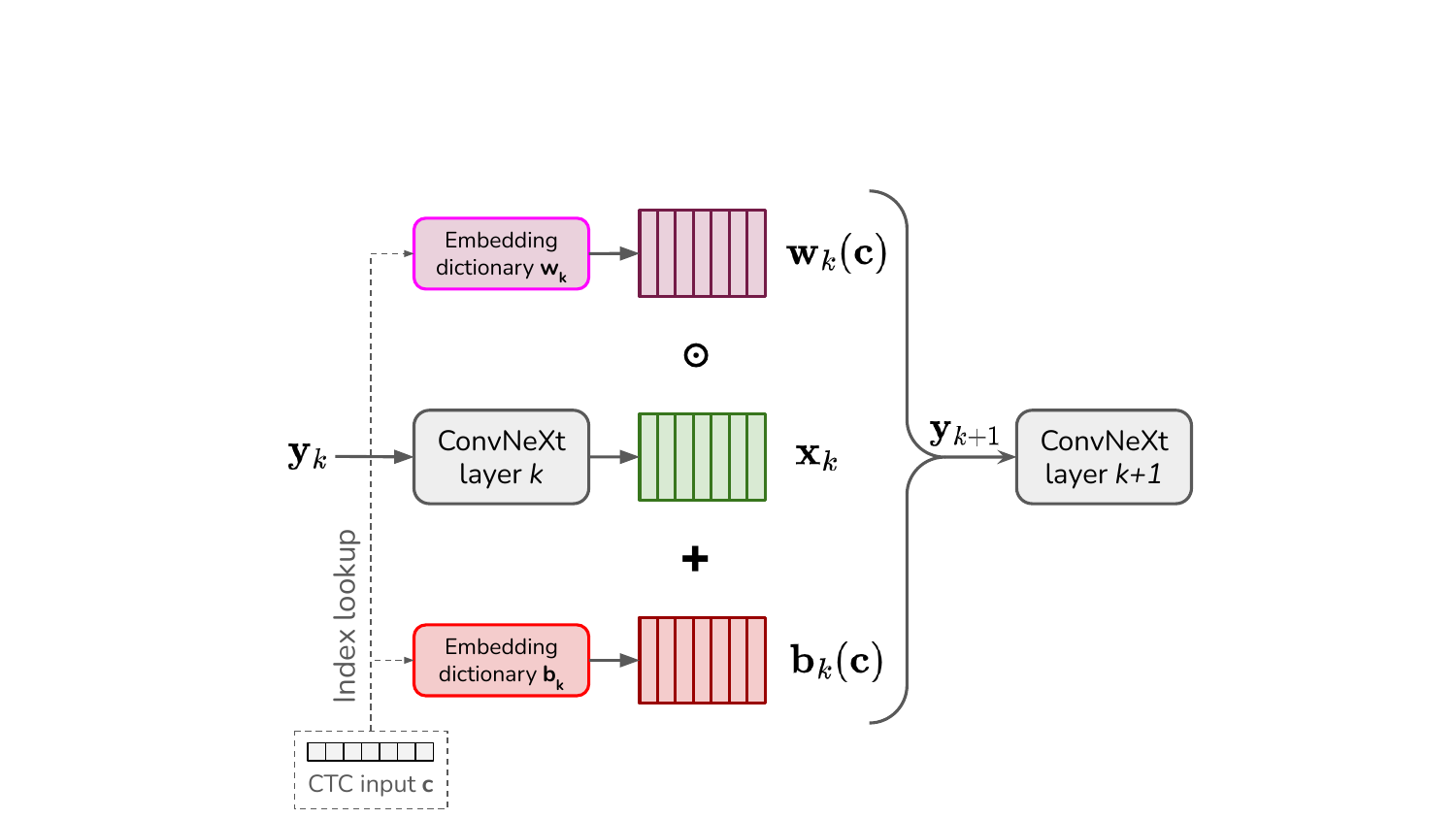}
    \label{fig:char_cond}
\end{figure}

As illustrated in Figure~\ref{fig:char_cond}, the $k$-th character conditioning layer (placed after the $k$-th ConvNeXt layer) takes $\ch$ as input and
contains two dictionaries of learnable embeddings both of dimension $D$, denoted  $\scaledict_k$ and $\biasdict_k$. 
Each dictionary has 31 entries, one for each CTC character.
We denote by $\scaledict_k(\ch) \in \mathbb{R}^{D\times T}$ the
matrix of embeddings constructed by
concatenating entries from $\scaledict_k$ according to the indexes in $\ch$.
Likewise, $\biasdict_k(\ch) \in \mathbb{R}^{D\times T}$ refers to the same operation applied to dictionary $\biasdict_k$.
Inspired by FiLM~\cite{perez_film_2018}, and given the output of the $k$-th ConvNeXt layer $\x_k$, we condition $\x_k$ with the affine transformation
\begin{equation}
    \label{eq:char_cond}
    \y_{k+1} = \scaledict_k(\ch) \odot \x_k + \biasdict_k(\ch)
\end{equation}
where $\odot$ represents element-wise multiplication. 
Each intermediate feature matrix in the vocoder is conditioned upon the time-aligned input character.
Output $\y_{k+1}$ is then used as the input to the following $(k+1)$-th ConvNeXt layer.
Note that, while the vocoder is conditioned by $\ch$, it is still driven by the acoustic tokens $\predicted$, meaning that the approach is still closer to voice conversion than text-to-speech. 


\subsection{Vocoder training}
The character embeddings in $\scaledict_k$ and $\biasdict_k$ are optimised jointly with the vocoder. 
The goal is to ensure high quality waveform synthesis and to preserve the spoken content contained within the input utterance $\utt$.
The first
goal is addressed using the training loss for the original Vocos vocoder~\cite{vocos}.  It comprises three components:
a mel-spectrogram reconstruction loss ($\ell_{\text{mel}}$); 
an adversarial loss against a set of multi-period and multi-resolution discriminators~\cite{kong_hifi-gan_2020, jang21_interspeech} ($\ell_{\text{gan}}$) in the style of a generative adversarial network (GAN)~\cite{the_one_and_only_gan_paper}; 
a feature-matching loss between the discriminators ($\ell_{\text{fm}}$).
These are represented by the two upper-most, red-coloured boxes in Figure~\ref{fig:train_inf}.
To encourage the preservation of spoken content, we add a new CTC loss component $\ell_{\text{ctc}}$ shown in the lower red box.
The training dataset is composed of triplets $(\utt_i, \ch_i, \Bar{\predicted}_{i})$, where $\utt_i$ is an unprotected utterance, $\ch_i = \text{ASR}_{\text{ctc}}\left(\utt_i\right)$ is the sequence of CTC symbols extracted from $\utt_i$ using $\text{ASR}_{\text{ctc}}$ and $\Bar{\predicted}_{i}$ is the set of acoustic tokens produced by the pair of transformers in the special case where the speaker prompt is that of the original speaker (no anonymisation).


During training, the vocoder receives as input $\Bar{\predicted}_{i}$ (for copy-synthesis in the usual way) and $\ch_i$ (for character-level conditioning) and generates the waveform $\Bar{\utt}_i$.
The CTC symbol sequence $\Bar{\ch}_i = \text{ASR}_{\text{ctc}}\left(\Bar{\utt}_i\right)$ is then inferred. The reconstruction objectives $\ell_{\text{mel}}$, $\ell_{\text{gan}}$ and $\ell_{\text{fm}}$ are computed by comparing $\utt_i$ and $\Bar{\utt}_i$, while $\ell_{\text{ctc}}$ is computed between $\ch_i$ and $\Bar{\ch}_i$.
The generator training loss is then
\begin{equation}
\ell_{\text{gen}} = \lambda_{\text{mel}}\ell_{\text{mel}} + \lambda_{\text{gan}}\ell_{\text{gan}} + \lambda_{\text{fm}}\ell_{\text{fm}} + \lambda_{\text{ctc}}\ell_{\text{ctc}}
\end{equation}
where hyperparameters $\lambda$ are weights assigned to each term.
The discriminator loss is the same as in~\cite{vocos}.
The generator (vocoder) and the discriminators are optimized with AdamW~\cite{adamw}.

\begin{table*}[ht]
\centering
\caption{Results on the evaluation sets of the VoicePrivacy Challenge 2024 of all baselines (B1 to B6) and our proposed system (highlighted in gray). The NAC system with character-level conditioning achieves the best results in terms of spoken content preservation (WER) and second best emotion preservation (UAR) while also maintaining a strong privacy protection level (EER).}
\begin{tblr}{Q[l,m] | Q[c,m] | Q[c,m] | Q[c,m] | Q[c,m] | Q[c,m] | Q[c,m] | Q[c,m] | Q[c,m,gray9] }
\hline[1.3pt]
\textbf{Dataset} & \textbf{Unprotected} & \textbf{B1} & \textbf{B2} & \textbf{B3} & {\textbf{B4}\\ \textbf{\footnotesize (original NAC)}} & \textbf{B5} & \textbf{B6} & \SetCell{gray8} {\textbf{NAC {\footnotesize + char.}} \\ \textbf{\footnotesize conditioning}} \\
\hline[1.3pt]
EER Libri-dev (\%)      & 5.72   & 9.20   & 7.48 & 25.24 & 32.71   & 34.37    & 23.05 & 30.76     \\
EER Libri-test (\%)     & 4,59   & 6.07   & 4.52 & 27.32 & 30.26   & 34.34    & 21.14 & 28.69     \\
\hline\hline
WER Libri-dev (\%)      & 1,80   & 3.07   & 10.44 & 4.29 & 6.15    & 4.74     & 9.69 & 2.66      \\
WER Libri-test (\%)     & 1,85   & 2.91   & 9.95 & 4.35 & 5.90    & 4.37     & 9.09 & 2.54      \\
\hline\hline
UAR IEMOCAP-dev (\%)    & 69.08  & 42.71  & 55.61 & 38.09 & 41.97   & 38.08    & 36.39 & 44.91     \\
UAR IEMOCAP-test (\%)   & 71.06  & 42.78  & 53.49 & 37.57 & 42.78   & 38.17    & 36.13 & 43.59     \\
\hline[1.3pt]
\end{tblr}
\label{tab:results}
\end{table*}

\section{Experimental setup}
\label{sec:setup}
The vocoder weights are initialised to those of the Vocos checkpoint trained using EnCodec 24~kHz tokens.\footnote{\url{https://huggingface.co/charactr/vocos-encodec-24khz}}
Both character embedding dictionaries $\scaledict_k$ and $\biasdict_k$ are initialised to $\mathbf{1}$ and $\mathbf{0}$ respectively, so that Eq.~\eqref{eq:char_cond} is equivalent to an identity operation.
The weights of the generator loss are $\lambda_{\text{mel}} = \lambda_{\text{fm}} = 1$, $\lambda_{\text{gan}} = 0.5$, $\lambda_{\text{ctc}} = 1.5$.
The full system is trained for 300k
steps with an initial learning rate of 5e-4 and with a single-cycle cosine annealing decreasing to 0.
Since the generator (the vocoder) weights are initialised to those of a pretrained checkpoint, we do the same for the discriminator since it was found to increase adversarial training stability. This is done by fixing the weights of the pretrained generator while optimising the discriminators for 6000 steps with an initial learning rate of \mbox{1e-3} and cosine annealing.
Both the generator and the discriminators are trained using the \emph{LibriSpeech-train-clean-100}~\cite{librispeech} database.
The $\text{ASR}_{\text{ctc}}$ system
is a pretrained SpeechBrain~\cite{speechbrain} model\footnote{\url{https://huggingface.co/speechbrain/asr-wav2vec2-librispeech}} and is a wav2vec 2.0 backbone with a classification head, both jointly fine-tuned using the LibriSpeech database.

Evaluation is performed using the 2024 VoicePrivacy Challenge protocol~\cite{vpc24}. 
Anonymisation is performed at the \emph{utterance level} using the voice of a pseudo-speaker selected at random from the speaker prompt pool.\footnote{Our inference code will be made available at \url{https://github.com/m-pana/spk_anon_nac_lm}.}
We use a semi-informed attacker scenario.  $\text{ASV}_{\text{eval}}^{\text{anon}}$ is an ECAPA-TDNN model~\cite{ecapa} trained using the \emph{LibriSpeech-train-clean-360} database similarly anonymised at the utterance level. 
$\text{ASR}_{\text{eval}}$ and $\text{SER}_{\text{eval}}$ models are pretrained using the original (unprotected) \emph{LibriSpeech-train-960} and \emph{IEMOCAP}~\cite{busso2008iemocap} databases, respectively (see~\cite{vpc24} for further details). 
EER, WER and UAR metrics are computed from anonymised data using the 2024 VoicePrivacy Challenge evaluation pipeline.\footnote{\url{https://github.com/Voice-Privacy-Challenge/Voice-Privacy-Challenge-2024}}

\section{Results}
\label{sec:results}
Results for the evaluation partition of the 2024 VoicePrivacy Challenge database are reported in Table~\ref{tab:results}. 
Also shown are results for the full set of the six challenge baselines.
Among them are:
B1, a simple x-vector--based anonymisation system for which the WER is the lowest among competing baselines~\cite{vpc24}; B4, the original NAC-based anonymisation system~\cite{nac} for which both the EER and UAR are competitive, but for which the WER is relatively high; B5, a vector quantisation-based system that achieves the highest EER~\cite{champion_thesis}.

With a WER of ${\sim}2.5\%$, our system outperforms all competing baselines in terms of preserving speech content, a relative improvement of almost 60\% over the original B4 system.
As an additional benefit, character-level conditioning also brings modest improvements to emotion preservation.  
UARs of 45\% and 44\% are also higher than those for all competing baselines. 
This improvement might be due to the encoding of some implicit prosody information in $\ch$.  
The information contained in $\ch$ is not exclusively `textual'. Since  $\ch$ associates a character to every time step, it can also encode nuances of the speaking rate which are pertinent to the SER task.

The benefits come at the expense of a modest loss in 
anonymisation performance. The EER falls from 
${\sim}31\%$ for the original B4 baseline to ${\sim}29\%$ with character conditioning.
The reason is likely the same as for the improvement in SER:
nuances of the speaking rate are also pertinent to the ASV task.
Nonetheless, 
anonymisation performance is still competitive.
Moreover, the relative decrease in the EER
of $6\%$ might be a small price to pay given the substantially greater relative improvement in the
WER of $56\%$.

\vspace{-0.1cm}
\section{Conclusions}
We propose character-level conditioning, a technique designed to preserve spoken content when 
a vocoder is employed for voice anonymisation.
An auxiliary connectionist temporal classification based automatic speech recognition model is used to extract a symbol sequence from the unprotected utterance. This is used to condition the vocoder module of the anonymisation system via a set of learnable embedding dictionaries in order to encourage the preservation of spoken content.
%
Relative to the baseline approach, and for only a modest cost in anonymisation performance, the technique
is successful in decreasing the word error rate 
computed from anonymised utterances 
by almost $60\%$. 
The resulting system outperforms all six of the 2024 VoicePrivacy challenge baselines in terms of preserving spoken content.
%
So long as gradients can be backpropagated, alternative approaches to the extraction of symbol sequences with appropriate temporal resolution could also be used.  The technique 
could 
equally be applied to other approaches to anonymisation which employ a
neural-based vocoder model for 
waveform synthesis.


\balance
\bibliographystyle{IEEEtran}
\bibliography{mybib}

\begin{thebibliography}{10}
\providecommand{\url}[1]{#1}
\csname url@samestyle\endcsname
\providecommand{\newblock}{\relax}
\providecommand{\bibinfo}[2]{#2}
\providecommand{\BIBentrySTDinterwordspacing}{\spaceskip=0pt\relax}
\providecommand{\BIBentryALTinterwordstretchfactor}{4}
\providecommand{\BIBentryALTinterwordspacing}{\spaceskip=\fontdimen2\font plus
\BIBentryALTinterwordstretchfactor\fontdimen3\font minus \fontdimen4\font\relax}
\providecommand{\BIBforeignlanguage}[2]{{%
\expandafter\ifx\csname l@#1\endcsname\relax
\typeout{** WARNING: IEEEtran.bst: No hyphenation pattern has been}%
\typeout{** loaded for the language `#1'. Using the pattern for}%
\typeout{** the default language instead.}%
\else
\language=\csname l@#1\endcsname
\fi
#2}}
\providecommand{\BIBdecl}{\relax}
\BIBdecl

\bibitem{literally_the_gdpr_itself}
\BIBentryALTinterwordspacing
{European Parliament} and {Council of the European Union}. Regulation ({EU}) 2016/679 of the {European} {Parliament} and of the {Council} of 27 {April} 2016 on the protection of natural persons with regard to the processing of personal data and on the free movement of such data, and repealing {Directive} 95/46/{EC} ({General} {Data} {Protection} {Regulation}). [Online]. Available: \url{https://data.europa.eu/eli/reg/2016/679/oj}
\BIBentrySTDinterwordspacing

\bibitem{nautsch19c_interspeech}
A.~Nautsch, C.~Jasserand, E.~Kindt, M.~Todisco, I.~Trancoso, and N.~Evans, ``{The GDPR \& Speech Data: Reflections of Legal and Technology Communities, First Steps Towards a Common Understanding},'' in \emph{Proc. Interspeech 2019}, 2019, pp. 3695--3699.

\bibitem{tomashenko_introducing_2020}
\BIBentryALTinterwordspacing
N.~Tomashenko, B.~M.~L. Srivastava, X.~Wang, E.~Vincent, A.~Nautsch, J.~Yamagishi, N.~Evans, J.~Patino, J.-F. Bonastre, P.-G. Noé, and M.~Todisco, ``Introducing the {VoicePrivacy} initiative,'' in \emph{Interspeech 2020}, pp. 1693--1697. [Online]. Available: \url{http://arxiv.org/abs/2005.01387}
\BIBentrySTDinterwordspacing

\bibitem{vpc24}
N.~Tomashenko, X.~Miao, P.~Champion, S.~Meyer, X.~Wang, E.~Vincent, M.~Panariello, N.~Evans, J.~Yamagishi, and M.~Todisco, ``{The VoicePrivacy 2024 Challenge Evaluation Plan},'' 2024.

\bibitem{vpc2020_results_and_findings}
\BIBentryALTinterwordspacing
N.~Tomashenko, X.~Wang, E.~Vincent, J.~Patino, B.~M.~L. Srivastava, P.-G. No\'{e}, A.~Nautsch, N.~Evans, J.~Yamagishi, B.~O’Brien, A.~Chanclu, J.-F. Bonastre, M.~Todisco, and M.~Maouche, ``{The VoicePrivacy 2020 Challenge: Results and findings},'' \emph{Comput. Speech Lang.}, vol.~74, no.~C, jul 2022. [Online]. Available: \url{https://doi.org/10.1016/j.csl.2022.101362}
\BIBentrySTDinterwordspacing

\bibitem{vpc22_perspectives}
M.~Panariello, N.~Tomashenko, X.~Wang, X.~Miao, P.~Champion, H.~Nourtel, M.~Todisco, N.~Evans, E.~Vincent, and J.~Yamagishi, ``{The VoicePrivacy 2022 Challenge: Progress and Perspectives in Voice Anonymisation},'' \emph{IEEE/ACM Transactions on Audio, Speech, and Language Processing}, vol.~32, pp. 3477--3491, 2024.

\bibitem{nac}
M.~Panariello, F.~Nespoli, M.~Todisco, and N.~Evans, ``Speaker anonymization using neural audio codec language models,'' in \emph{ICASSP 2024 - 2024 IEEE International Conference on Acoustics, Speech and Signal Processing (ICASSP)}, 2024, pp. 4725--4729.

\bibitem{deng_v-cloak_2023}
\BIBentryALTinterwordspacing
J.~Deng, F.~Teng, Y.~Chen, X.~Chen, Z.~Wang, and W.~Xu, ``{V-Cloak: Intelligibility-, Naturalness- \& Timbre-Preserving Real-Time Voice Anonymization},'' pp. 5181--5198. [Online]. Available: \url{https://www.usenix.org/conference/usenixsecurity23/presentation/deng-jiangyi-v-cloak}
\BIBentrySTDinterwordspacing

\bibitem{ctc}
\BIBentryALTinterwordspacing
A.~Graves, S.~Fern\'{a}ndez, F.~Gomez, and J.~Schmidhuber, ``Connectionist temporal classification: labelling unsegmented sequence data with recurrent neural networks,'' in \emph{Proceedings of the 23rd International Conference on Machine Learning}, ser. ICML '06.\hskip 1em plus 0.5em minus 0.4em\relax New York, NY, USA: Association for Computing Machinery, 2006, p. 369–376. [Online]. Available: \url{https://doi.org/10.1145/1143844.1143891}
\BIBentrySTDinterwordspacing

\bibitem{T04}
S.~Meyer, P.~Tilli, F.~Lux, P.~Denisov, J.~Koch, and N.~T. Vu, ``{Cascade of phonetic speech recognition, speaker embeddings gan and multispeaker speech synthesis for the VoicePrivacy 2022 Challenge },'' in \emph{Proc. 2nd Symposium on Security and Privacy in Speech Communication}, 2022.

\bibitem{meyer_prosody_is_not_identity}
S.~Meyer, F.~Lux, J.~Koch, P.~Denisov, P.~Tilli, and N.~T. Vu, ``Prosody is not identity: A speaker anonymization approach using prosody cloning,'' in \emph{ICASSP 2023 - 2023 IEEE International Conference on Acoustics, Speech and Signal Processing (ICASSP)}, 2023, pp. 1--5.

\bibitem{ohnn}
X.~Miao, X.~Wang, E.~Cooper, J.~Yamagishi, and N.~Tomashenko, ``Speaker anonymization using orthogonal householder neural network,'' \emph{IEEE/ACM Transactions on Audio, Speech, and Language Processing}, vol.~31, pp. 3681--3695, 2023.

\bibitem{vpc2022}
N.~Tomashenko, X.~Wang, X.~Miao, H.~Nourtel, P.~Champion, M.~Todisco, E.~Vincent, N.~Evans, J.~Yamagishi, and J.-F. Bonastre, ``{The VoicePrivacy 2022 Challenge Evaluation Plan},'' 2022.

\bibitem{encodec}
\BIBentryALTinterwordspacing
A.~D{\'e}fossez, J.~Copet, G.~Synnaeve, and Y.~Adi, ``High fidelity neural audio compression,'' \emph{Transactions on Machine Learning Research}, 2023, featured Certification, Reproducibility Certification. [Online]. Available: \url{https://openreview.net/forum?id=ivCd8z8zR2}
\BIBentrySTDinterwordspacing

\bibitem{vocos}
\BIBentryALTinterwordspacing
H.~Siuzdak, ``Vocos: Closing the gap between time-domain and fourier-based neural vocoders for high-quality audio synthesis,'' in \emph{The Twelfth International Conference on Learning Representations}, 2024. [Online]. Available: \url{https://openreview.net/forum?id=vY9nzQmQBw}
\BIBentrySTDinterwordspacing

\bibitem{convnext}
\BIBentryALTinterwordspacing
Z.~Liu, H.~Mao, C.-Y. Wu, C.~Feichtenhofer, T.~Darrell, and S.~Xie, ``A {ConvNet} for the 2020s,'' pp. 11\,976--11\,986. [Online]. Available: \url{https://openaccess.thecvf.com/content/CVPR2022/html/Liu_A_ConvNet_for_the_2020s_CVPR_2022_paper.html}
\BIBentrySTDinterwordspacing

\bibitem{perez_film_2018}
E.~Perez, F.~Strub, H.~de~Vries, V.~Dumoulin, and A.~Courville, ``{FiLM}: visual reasoning with a general conditioning layer,'' in \emph{Proceedings of the Thirty-Second {AAAI} Conference on Artificial Intelligence and Thirtieth Innovative Applications of Artificial Intelligence Conference and Eighth {AAAI} Symposium on Educational Advances in Artificial Intelligence}, ser. {AAAI}'18/{IAAI}'18/{EAAI}'18.\hskip 1em plus 0.5em minus 0.4em\relax {AAAI} Press, pp. 3942--3951.

\bibitem{kong_hifi-gan_2020}
\BIBentryALTinterwordspacing
J.~Kong, J.~Kim, and J.~Bae, ``{HiFi}-{GAN}: Generative adversarial networks for efficient and high fidelity speech synthesis,'' in \emph{Advances in Neural Information Processing Systems}, vol.~33.\hskip 1em plus 0.5em minus 0.4em\relax Curran Associates, Inc., pp. 17\,022--17\,033. [Online]. Available: \url{https://proceedings.neurips.cc/paper/2020/hash/c5d736809766d46260d816d8dbc9eb44-Abstract.html}
\BIBentrySTDinterwordspacing

\bibitem{jang21_interspeech}
W.~Jang, D.~Lim, J.~Yoon, B.~Kim, and J.~Kim, ``{UnivNet: A Neural Vocoder with Multi-Resolution Spectrogram Discriminators for High-Fidelity Waveform Generation},'' in \emph{Proc. Interspeech 2021}, 2021, pp. 2207--2211.

\bibitem{the_one_and_only_gan_paper}
\BIBentryALTinterwordspacing
I.~Goodfellow, J.~Pouget-Abadie, M.~Mirza, B.~Xu, D.~Warde-Farley, S.~Ozair, A.~Courville, and Y.~Bengio, ``Generative adversarial networks,'' \emph{Commun. ACM}, vol.~63, no.~11, p. 139–144, oct 2020. [Online]. Available: \url{https://doi.org/10.1145/3422622}
\BIBentrySTDinterwordspacing

\bibitem{adamw}
\BIBentryALTinterwordspacing
I.~Loshchilov and F.~Hutter, ``Decoupled weight decay regularization,'' in \emph{International Conference on Learning Representations}, 2019. [Online]. Available: \url{https://openreview.net/forum?id=Bkg6RiCqY7}
\BIBentrySTDinterwordspacing

\bibitem{librispeech}
V.~Panayotov, G.~Chen, D.~Povey, and S.~Khudanpur, ``{Librispeech: An ASR corpus based on public domain audio books},'' in \emph{2015 IEEE International Conference on Acoustics, Speech and Signal Processing (ICASSP)}, 2015, pp. 5206--5210.

\bibitem{speechbrain}
M.~Ravanelli, T.~Parcollet, P.~Plantinga, A.~Rouhe, S.~Cornell, L.~Lugosch, C.~Subakan, N.~Dawalatabad, A.~Heba, J.~Zhong, J.-C. Chou, S.-L. Yeh, S.-W. Fu, C.-F. Liao, E.~Rastorgueva, F.~Grondin, W.~Aris, H.~Na, Y.~Gao, R.~D. Mori, and Y.~Bengio, ``{SpeechBrain}: A general-purpose speech toolkit,'' 2021, arXiv:2106.04624.

\bibitem{ecapa}
\BIBentryALTinterwordspacing
B.~Desplanques, J.~Thienpondt, and K.~Demuynck, ``{ECAPA}-{TDNN}: Emphasized channel attention, propagation and aggregation in {TDNN} based speaker verification,'' in \emph{Interspeech 2020}.\hskip 1em plus 0.5em minus 0.4em\relax {ISCA}, pp. 3830--3834. [Online]. Available: \url{https://www.isca-speech.org/archive/interspeech_2020/desplanques20_interspeech.html}
\BIBentrySTDinterwordspacing

\bibitem{busso2008iemocap}
C.~Busso, M.~Bulut, C.-C. Lee, A.~Kazemzadeh, E.~Mower, S.~Kim, J.~N. Chang, S.~Lee, and S.~S. Narayanan, ``Iemocap: Interactive emotional dyadic motion capture database,'' \emph{Language resources and evaluation}, vol.~42, pp. 335--359, 2008.

\bibitem{champion_thesis}
P.~Champion, ``Anonymizing speech: Evaluating and designing speaker anonymization techniques,'' Ph.D. dissertation, Université de Lorraine, 2023.

\end{thebibliography}

\end{document}